\documentclass[aps,amssymb,amsmath,showpacs,twocolumn,tightenlines]{revtex4}
\usepackage{latexsym}
\input epsf

\begin{document}
\title{Event Texture Search for Phase Transitions in $Pb+Pb$ Collisions}
\author{
The NA44 Collaboration \\
I.~Bearden$^{a}$,
 H.~B$\rm{\not\!{o}}$ggild$^{a}$,
J.~Boissevain$^{b}$,
L.~Conin$^{d}$,
J.~Dodd$^{c}$,
B.~Erazmus$^{d}$,
S.~Esumi$^{e}$\footnotemark[1]\footnotetext[1]{
now at KEK -- High Energy Accelerator Research Organization,
1-1 Oho, Tsukuba, Ibaraki 305,  Japan
},
C.~W.~Fabjan$^{f}$,
D.~Ferenc$^{g}$,
D.~E.~Fields$^{b}$
\footnotemark[2]\footnotetext[2]{
now at University of New Mexico, Albuquerque, NM 87131, USA},
A.~Franz$^{f}$ \footnotemark[3]
\footnotetext[3]{now at Brookhaven National Laboratory, Upton,
 NY 11973, USA},
J.~J.~Gaardh$\rm{\not\!{o}}$je$^{a}$,
A.~G.~Hansen$^{a}$\footnotemark[11]\footnotetext[11]{
now at Los Alamos National Laboratory, Los Alamos, NM 87545, USA
},
O.~Hansen$^{a}$,
D.~Hardtke$^{i}$\footnotemark[4]\footnotetext[4]{
now at Lawrence Berkeley National Laboratory,
              Berkeley, CA 94720, USA
},
H.~van~Hecke$^{b}$,
E.~B.~Holzer$^{f}$,
T.~J.~Humanic$^i$,
P.~Hummel$^f$,
B.~V.~Jacak$^j$,
R.~Jayanti$^i$,
K.~Kaimi$^e$ \footnotemark[5]\footnotetext[5]{deceased},
M.~Kaneta$^e$,
T.~Kohama$^e$,
M.~L.~Kopytine$^j$\footnotemark[6]\footnotetext[6]{
on an unpaid leave from P. N. Lebedev Physical Institute,
Russian Academy of Sciences},
M.~Leltchouk$^c$,
A.~Ljubicic, Jr$^g$,
B.~L{\"o}rstad$^k$,
N.~Maeda$^e$ \footnotemark[7]\footnotetext[7]{
now at Florida State University, Tallahassee, FL 32306, USA},
L.~Martin$^d$,
A.~Medvedev$^c$,
M.~Murray$^h$,
H.~Ohnishi$^e$ \footnotemark[3],
G.~Paic$^f$
\footnotemark[8]\footnotetext[8]
{affiliated with Ohio State University, Columbus, OH 43210, USA},
S.~U.~Pandey$^i$,
F.~Piuz$^f$,
J.~Pluta$^d$ \footnotemark[9]\footnotetext[9]
{Institute of Physics, Warsaw University of Technology,
  Koszykowa 75, 00-662 Warsaw, Poland},
V.~Polychronakos$^l$,
M.~Potekhin$^c$,
G.~Poulard$^f$,
D.~Reichhold$^i$,
A.~Sakaguchi$^e$\footnotemark[10]\footnotetext[10]
{now at Osaka University, Toyon
aka, Osaka 560-0043,
Japan},
J.~Schmidt-S$\rm{\not\!{o}}$rensen$^k$,
J.~Simon-Gillo$^b$,
W.~Sondheim$^b$,
T.~Sugitate$^e$,
J.~P.~Sullivan$^b$,
Y.~Sumi$^e$,
W.~J.~Willis$^c$,
K.~L.~Wolf$^h$ \footnotemark[5],
N.~Xu$^b$ \footnotemark[4],
D.~S.~Zachary$^i$}
\address{
$^a$ Niels Bohr Institute,DK-2100, Copenhagen, Denmark;\\
$^b$ Los Alamos National Laboratory, Los Alamos, NM 87545, USA;\\
$^c$ Columbia University, New York, NY 10027, USA;\\
$^d$ Nuclear Physics Laboratory of Nantes, 44072 Nantes, France;\\
$^e$ Hiroshima University, Higashi-Hiroshima 739, Japan;\\
$^f$ CERN, CH-1211 Geneva 23, Switzerland;\\
$^g$ Rudjer Boscovic Institute, Zagreb, Croatia;\\
$^h$ Texas A\&M University, College Station, Texas 77843, USA;\\
$^i$ The Ohio State University, Columbus, OH 43210, USA;\\
$^j$ SUNY at Stony Brook, Stony Brook, NY 11794, USA;\\
$^k$ University of Lund, S-22362 Lund, Sweden;\\
$^l$ Brookhaven National Laboratory, Upton, NY 11973, USA.\\
}

\date{October 1, 2001}

\begin{abstract}
	NA44 uses a 512 channel Si pad array covering
	$1.5 <\eta < 3.3$
	to study charged hadron production in 158 A GeV $Pb+Pb$
	collisions at 	the CERN SPS.
	We apply a multiresolution analysis, based on a Discrete 
	Wavelet Transformation, to probe the texture of particle 
	distributions 	event-by-event, allowing
	simultaneous localization of features in space and scale.
	Scanning a
	broad range of multiplicities, we search for
        signals of clustering and of critical
	behavior in the power spectra of local density fluctuations.
	The data
	are compared with detailed simulations of detector response, 
	using
	heavy ion event generators, and with a reference sample
	created via event mixing.
	An upper limit is set on the probability and magnitude
	of dynamical fluctuations.
\end{abstract}
\maketitle

\section{Introduction}

The main experimental challenge in relativistic heavy ion collisions
is to find evidence for the expected
QCD phase transition at high temperature.
Deconfinement and chiral symmetry restoration are expected to take place
during the hot, strongly interacting stage early in the collision.
As a phase transition in such collisions is inherently
 a multiparticle phenomenon,
multiparticle observables, defined on event-by-event basis,
are of great interest.
Recently published event-by-event analyses of the 158 GeV/A $Pb+Pb$ data
either analyze a small number of events \cite{EMU15} in great detail,
or analyze properties of a large ensemble of events
using a single observable ($p_T$)
to compare different ensemble averages\cite{NA49_phi}.
In the first case,
accumulation of feature information from large data sets
remains an open issue.
In the second case, an ensemble average
on a set of \emph{post-freeze-out} events is not representative of the
\emph{pre-freeze-out} history of those events,
due to the dramatic non-stationarity of the open system, with a 
consequent lack of ergodicity.
Violations of ergodicity generally happen in the course of phase
transitions\cite{broken_ergo}.

We concentrate on \emph{texture}, or \emph{local
fluctuation} observables,
analyzing single events independently to determine
the scale composition of fluctuations.
In the following, we may omit the term ``local'',
but we will always talk about fluctuations in the particle density from
one point to another within a single event, i.e. in the \emph{local} 
sense, as opposed to fluctuations of \emph{global} quantities from 
one event to another.

In 1985, L. Van Hove formulated a model of quark-gluon plasma
hadronization\cite{VanHove:1985zy} with a first order phase transition.
Longitudinal expansion of the colliding system, with particle formation via
string or color flux tube breaking, can result in plasma droplets as
large as a few fm across. 
The droplets hadronize by \emph{deflagration}\cite{deflagration}.
This is expected to result in $\,dN/\,dy$ distributions with bumps or spikes
on top of an otherwise smooth structure. 
Other models\cite{Mardor_Svetitsky} also predict 
bubbles of one phase embedded in the other.

In the absence of a direct, event-by-event observable-based test of 
these
predictions, the picture had been further developed
\cite{Cleymans_93,Cleymans_94}
in order to connect it
with the traditional observables such as the $m_T$ slope parameter $T$
and the baryon and strangeness chemical potentials:
the hadron ``temperatures'' $T$ in the SPS data are
 higher than lattice QCD predictions for a phase transition
temperature.
Using a first order phase transition hydrodynamical
model with a sharp front between the phases, Bilic \emph{et al.}
\cite{Cleymans_93,Cleymans_94}
concluded that a QGP \emph{supercooling} and hadron gas
 \emph{superheating}
is a consequence of the continuity equations and of the requirement that
the entropy be increased in the transition.
In the case of bubbles in the QGP phase, the plasma deflagrates; 
otherwise, it detonates.
A direct measurement of the hadron texture at freeze-out, if it
detects presence of the droplets/bubbles, could provide an argument in
favor of the first order phase transition. 


     The order of the confinement phase transition is still under
debate. It is a fluctuation driven first order transition 
\cite{Svetitsky_Yaffe,Pisarski_Wilczek}
in SU(3) with three massless quarks, but second order in the
case of finite mass \cite{FRBrown}
 or infinitely massive \cite{Svetitsky_Yaffe,Pisarski_Wilczek} 
strange quarks. 
A tricitical point may exist, separating the first order
transition from a second order transition with the same critical
exponents as the 3D Ising model \cite{Svetitsky_Yaffe}. 
For a second order phase
transition, local fluctuations of isospin or enhanced correlation
lengths may be observable \cite{Rajagopal_Wilczek,tricritical}.
 Large scale correlations
formed early in the collision are more likely to survive
diffusion in the later stages. Small scale fluctuations, on the
contrary, are more easily washed out by diffusion due to secondary
scattering among the hadrons \cite{Shuryak:2001pd}. 
Consequently, an analysis
method which can identify fluctuations on any scale is desirable.
In this paper, we utilize a Discrete Wavelet Transformation, which
has this property.

A wavelet is a function, 0 everywhere except for a well localized
spot.
For a pad detector, the discrete positions of the spots correspond naturally
to the pad positions, and the possible scales are multiples of the pad sizes. 
The scale is an analog of a Fourier frequency.
Location has no analog in the Fourier transform, and it provides an 
additional degree of analytical power, which explains much of the success
that wavelets met in the field of data processing and pattern recognition. 
(Examples of Fourier-based analyses of large scale azimuthal texture
in the field of relativistic heavy
ion collisions exist as well \cite{flow_AGS,NA49_elliptic}; this is how
the elliptic flow at relativistic energies was measured.) 
The binning of charged particle density inherent in measurements with
a segmented detector such as a Si pad detector makes the Haar wavelet a natural
choice of analyzing function; a Haar wavelet is a step function with given 
width, oscillating around zero with a single period. 

Discrete Wavelet Transformation (DWT)\cite{DWT}  
quantifies contributions of different  $\phi$ and $\eta$ scales
to the  event texture.
We use DWT to test for possible
large scale enhancement, as a function of the collision centrality.
We report the DWT
power spectrum  in pseudorapidity
$\eta$ and azimuthal angle $\phi$, for different charged particle
multiplicities.
We use mixed events to remove trivial fluctuations and background effects.

\section{Experimental setup}
\label{Section:setup}

\begin{figure}
\epsfxsize=8.6cm
\epsfbox{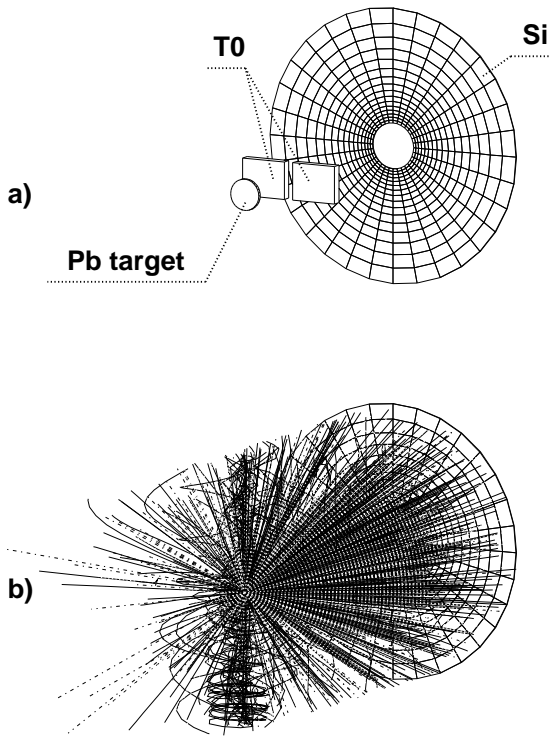}
\caption{
 a) The experimental setup: the target, the Si pad array and the T0
 scintillation counter. See text for a description of the
 detectors.
 b) The setup exposed to an RQMD event (GEANT simulation).
Magnetic field is on.
}
\label{setup}
\end{figure}
The experimental setup \cite{NA44ex} is shown in Fig.\ref{setup}.
The NA44 Si pad array,
installed 10 cm downstream from the target, in the magnetic field of the
first dipole,
measured ionization energy loss of charged particles
 in its 512 300 $\mu$m thick Si pads.
The plastic scintillator T0 (two rectangles seen in Fig.\ref{setup})
was used for a centrality trigger.
The SPS beam was collimated to a $1\times2$ mm profile.
T0 covered $ 1.4 \le \eta \le 3.7 $  for an $\eta$-dependent
fraction of azimuthal angle,
$0.22 \le \Delta \phi / 2\pi \le 0.84 $ respectively.
The silicon detector had
inner radius 7.2 mm and outer radius 43 mm, covering 
$1.5 \le \eta \le 3.3$.
The detector
was split \emph{radially} into 16
rings of equal
$\eta$ coverage.
Each  ring was further divided
\emph{azimuthally} into 32 sectors of equal angular coverage to form 
pads. The pads were read out by AMPLEX
\cite{AMPLEX}
chips, one chip per sector.
$\delta$-electrons, produced by the $Pb$ beam traversing
the target, were swept away to one side by the dipole magnetic field
($\le 1.6$ T).
Only the $\delta$-electron-free side
was used in this analysis.
Only 4 of the remaining 256 channels were inoperative.

An amplitude distribution from a typical channel, observed in the 
physics run and digitized with a 256 channel ADC is shown on 
Fig.\ref{si_adc}.
Channel pedestals had, on the average,  $FWHM=0.48 <dE>$
of 1 MIP.
In the texture analysis, every event was represented by a 2D array of
the calibrated digitized amplitudes of the channels ( an
\emph{amplitude array}).
Empty target runs were used to measure the background, and
cross-talk in the detector was evaluated off-line.


\section{Analysis technique}
\label{Section:analysis}
\subsection{Detector calibration}
The NA44 spectrometer information was not used in this analysis,
which focussed on the Si pad array data.

\begin{figure}
\epsfxsize=8.6cm
\epsfbox{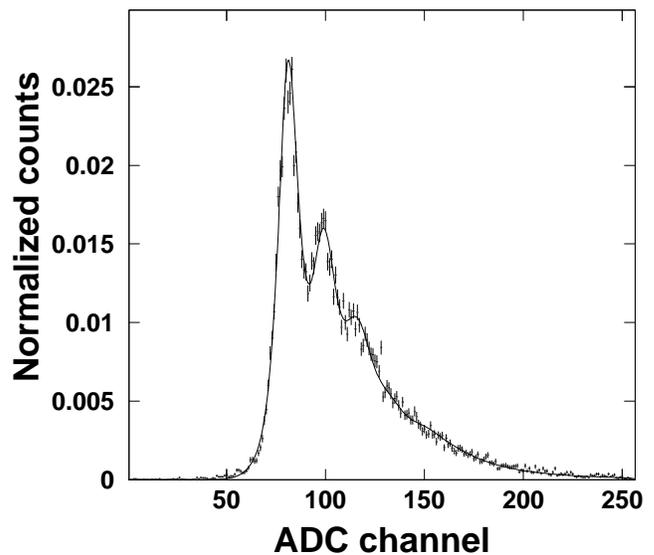}
\caption{Digitized amplitude distribution from channel 1 of the Si pad
array. The smooth curve shows a minimum $\chi^2$ Landau fit performed 
in course of the amplitude calibration.
The pedestal, the single and double hit peaks are distinguishable. }
\label{si_adc}
\end{figure}

ADC pedestals were fitted channel by channel
with a realistic functional shape, determined from low multiplicity
events in a minimum bias triggered run.
Amplitude calibration  of the Si detector was carried out channel 
by channel,
by fitting the amplitude distribution with a sum of single, double,
triple,
\emph{etc.} (up to septuple)  minimum ionizing particle Landau
distributions\cite{Landau_distr} with variable weights.
The Landau distributions were numerically convoluted with the pedestal
shape to  account for noise in the fit.
A typical fit from a single channel is shown in Fig.~\ref{si_adc}.
Parameters of the fit were used to simulate noise in a GEANT-based
detector response Monte Carlo code.

An offset of the event vertex with respect to the detector's symmetry 
axis results in a non-trivial functional dependence between the actual
$\eta$ and
$\phi$,
and the $\eta^\prime$,
$\phi^\prime$
presumed based on
the ``ideal'' geometry:
$\eta = \eta(\eta^\prime,\phi^\prime)$,
$\phi = \phi(\eta^\prime,\phi^\prime)$.
This makes the observable multiplicity distribution
$\,d^2N/\,d\phi^\prime\,d\eta^\prime$
(in the presumed coordinates) differ from a simple function of 
$\eta^\prime$:
\begin{equation}
\frac{\,d^2N}{\,d\phi^\prime\,d\eta^\prime}
 \not= \frac{1}{2\pi} \frac{\,dN}{\,d\eta^\prime}
\label{Jacobian_effect}
\end{equation}
In  the true coordinates $\eta$ and $\phi$,
the inequality \ref{Jacobian_effect} becomes an equality.
However, the detector's acceptance area in the true coordinates becomes
distorted.
In the following we will refer to this as a ``Jacobian effect''.
The Jacobian effect, obviously, contributes to the event textures,
especially on the large scale, and needs to be evaluated and corrected
for.

From Eq.\ref{Jacobian_effect}, the criterion of the true coordinate
basis $(\eta,\phi)$ emerges naturally: it is the basis which
makes the observable $\,d^2N/\,d\phi\,d\eta$ independent of $\phi$.
The minimization problem was solved numerically with 
MINUIT\cite{MINUIT},
and the resulting offsets are within the tolerance of the detector/beam
position.
Cross-talk between  the electronics channels is a detector-related
 correlation phenomenon and introduces a  ``texture'' effect of its own.
Both global cross-talk in the AMPLEX read-out chip \cite{AMPLEX}
and read-out board cross-talk are expected.
In our detector with 512 channels, there are $512\times(512-1)/2=130816$
two channel pairs (unordered),
all of which were subjected to covariance analysis off-line.
To magnify the non-trivial instrumental
contribution to the covariance matrix elements, we analyzed covariances
not between the amplitudes $A_i$ of channels $i$
themselves, but between
\begin{equation}
a_i =
A_i - \frac{\sum_{\mbox{half-ring of } i} A_k}
{\sum_{\mbox{half-ring of } i}1}
= A_i - \frac{1}{16}\sum_{\mbox{half-ring of } i} A_k
\end{equation}
Otherwise, the dominant contributor to the cov($A_i$,$A_j$) is
the trivial variation of the event's common multiplicity
\footnote{The negative autocorrelation introduced
thereby between certain channels is immaterial since our goal
is just one effective cross-talk coupling parameter for the detector.}.
Using this method, we concluded that the effective cross-talk coupling
was non-negligible only for neighboring channels within the same chip;
it was found to be 8.5\%.
As a remedy,
a chip-wise (i.e. sector-wise)
event mixing technique
including cross-talk in the reference sample
was used to construct a reference event sample.

\begin{figure}[!th]
\epsfxsize=8.6cm
\epsfbox{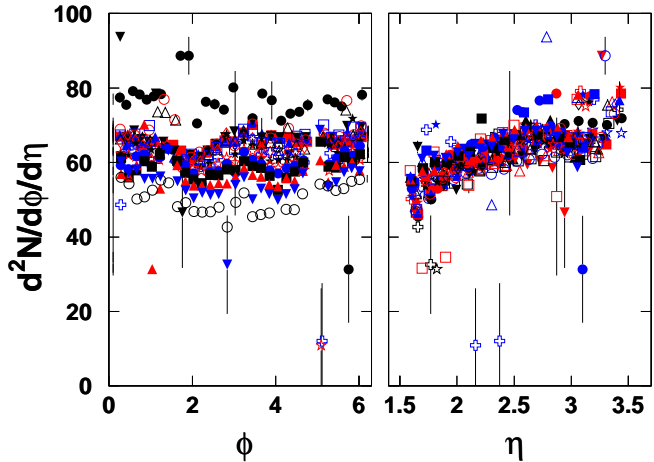}
\caption{
Double differential multiplicity distributions of charged particles
plotted as a function of azimuthal angle $\phi$
(with different symbols representing different rings)
and of pseudorapidity $\eta$
(with different symbols representing different sectors).
The $\phi$ and $\eta$ are in the \emph{aligned} coordinates.
 }
\label{d2n_deta_dphi}
\end{figure}
The double differential multiplicity data (Fig. \ref{d2n_deta_dphi})
illustrate the quality
of the detector operation,
calibrations,
geometrical alignment and Jacobian correction.
The data set is composed of two pieces, obtained by switching
the magnetic field polarity:
a negative polarity run
is used for sectors 9 to 24 (range of $\pi/2<\phi<3\pi/2$);
a positive polarity run
is used for sectors 1 to 8 and 25 to 32
(range of $0<\phi<\pi/2$ and $3\pi/2<\phi<2\pi$).
The reason to disregard one side of the detector is additional
occupancy due to $\delta$-electrons,
as was explained in section
\ref{Section:setup}.
Figure \ref{d2n_deta_dphi} demonstrates the quality of alignment as 
well,  since the $\eta$
and $\phi$ along the horizontal axes are the aligned coordinates.
Any geometrical offset of the detector makes acceptances of different
pads non-equal
and dependent on the pad position.
The acceptance of each pad has been calculated in the aligned 
coordinates,
and the $\,d^2N/\,d\phi\,d\eta$ uses the actual acceptances $\,d\phi$.
The shape of the $\phi$ dependence of $\,d^2N/\,d\phi\,d\eta$
(left panel of Fig. \ref{d2n_deta_dphi}) is flat
as it should be for an event ensemble with no reaction plane selection.
The $\eta$ dependence (right panel of Fig. \ref{d2n_deta_dphi}) shows
increasing  multiplicity towards midrapidity,
as is expected.
As can be seen from Fig.\ref{d2n_deta_dphi},
the detector's acceptance is asymmetric around midrapidity.
A correction for the cross-talk has been applied.

\subsection{Discrete Wavelet Transformation (DWT)}

Discrete wavelets are a set of functions, each having a proper width,
or scale, and a proper location so that the function differs from 0
only within that width and around that location.
The set of possible scales and locations is discrete.
The DWT formalizes the two dimensional particle distribution
in each $Pb+Pb$ collision  in pseudorapidity $\eta$ and azimuthal
angle $\phi$ by performing an image analysis -- transforming the event
 into a set of functions orthogonal with
respect to scale and location in  the ($\eta$, $\phi$) space.
We accumulate texture information  
by averaging the power spectra of many events.

The simplest DWT basis is the Haar wavelet, built upon the \emph{scaling 
function}\footnote{Some authors call it  ``mother function''.}
$g(x) = 1$ for $0\le x<1$ and 0 otherwise.
The function
\begin{equation}
 f(x) = \left\{ \begin{array}
       {r@{\quad:\quad}l}
      +1 & 0\le x<\frac{1}{2} \\ -1 & \frac{1}{2}\le x<1 \\ 0 & 
      otherwise
      \end{array} \right.
\end{equation}
is the wavelet function\footnote{Some authors call it ``father function''.}.

If the interaction vertex lies on the detector's symmetry axis,
every pad's acceptance is a rectangle in the $(\phi,\eta)$ space.
Then, the Haar basis is the natural choice, as its scaling function in
two dimensions (2D)
$G(\phi,\eta) = g(\phi)g(\eta)$
is just a pad's acceptance (modulo units).
We  set up a two dimensional (2D) wavelet basis:
\begin{equation}
F^{\lambda}_{m,i,j}(\phi,\eta) =
 2^{m}F^{\lambda}(2^{m}\phi-i,2^{m}\eta-j).
\label{wavelet_2D}
\end{equation}
The scaling function in 2D is
$G(\phi,\eta)=g(\phi)g(\eta)$.
As in Eq.\ref{wavelet_2D}, we construct $G_{m,i,j}(\phi,\eta)$ where $m$ is the integer
scale fineness index, and $i$ and $j$ index the positions of bin centers in 
$\phi$ and $\eta$
($1 \le m \le 4$ and $1\le i,j \le 16$
because we use $16=2^4$ rings and 16 sectors).
Different values of $\lambda$
(denoted as $\phi$, $\eta$, and $\phi\eta$) distinguish, respectively,
functions with  azimuthal, pseudorapidity, and diagonal texture
sensitivity:
\begin{equation}
F^\phi=f(\phi)g(\eta), \ \
F^\eta=g(\phi)f(\eta), \ \
F^{\phi\eta}=f(\phi)f(\eta)
\end{equation}
Then, $F^\lambda_{m,i,j}$ with integer $m$, $i$, and $j$ are known
\cite{DWT}
to form an orthonormal basis in the space
of all \emph{measurable functions} defined on the continuum of real
numbers $L^2({\mathbb{R}})$.
\begin{figure}
\epsfxsize=8.6cm
\epsfbox{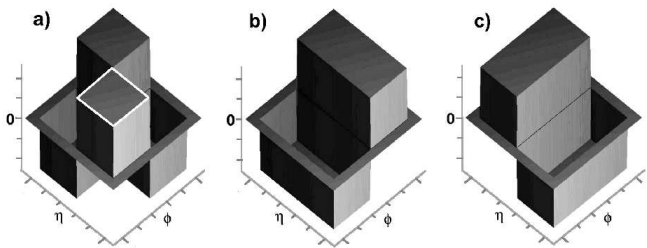}
\caption{Haar wavelet basis in two dimensions. The three modes of directional sensitivity are:
a) diagonal b) azimuthal c) pseudorapidity. For the finest scale used, the acceptance of a Si
pad would correspond to the white rectangle drawn ``on top'' of the function in panel a).
Every subsequent coarser scale is obtained by expanding
the functions of the previous scale by a 
factor of 2 in both dimensions. 
}
\label{haar}
\end{figure}
Fig. \ref{haar} shows the wavelet basis functions $F$ in two dimensions.
At first glance it might seem surprising that, unlike the 1D case, both $f$ and
$g$ enter the wavelet basis in 2D.
Fig. \ref{haar} clarifies this: in order to fully encode an arbitrary
shape of a measurable 2D function, one considers it as an addition of a
change along $\phi$ ($f(\phi)g(\eta)$, panel (b)), 
a change along $\eta$ ($g(\phi)f(\eta)$, panel (c)), and
a saddle-point pattern ($f(\phi)f(\eta)$, panel (c)), 
added with appropriate weight (positive, negative or zero), for a variety
of scales.
The finest scale available is determined by the detector segmentation, while
the coarser scales correspond to successively rebinning the track distribution.
The analysis is best visualized by considering the scaling function
$G_{m,i,j}(\phi,\eta)$ as binning the track distribution 
$\rho(\phi,\eta)$
in bins $i$,$j$
of fineness $m$, while the set of wavelet functions 
$F^{\lambda}_{m,i,j}(\phi,\eta)$ (or, to be exact, the wavelet expansion
 coefficients $\langle \rho, F^{\lambda}_{m,i,j}\rangle$)
gives the difference distribution between the data binned with given 
coarseness and that with binning one step finer.

While
the DWT analyzes the object  (an image, a sequence of data points, a
data array) by transforming it,
the \emph{full} information content inherent in the object is preserved
in the transformation.


We adopt the existing \cite{DWT_power} 1D DWT power spectrum analysis
technique  and expand it to 2D.
The track density in an individual event is
$\rho(\phi,\eta)$ and its \emph{local} fluctuation
in a given event is
$ \sigma^2 \equiv \langle \rho - \bar{\rho},\rho - \bar{\rho}\rangle,$
where $\bar{\rho}$ is the average $\rho$ (over the acceptance)
in the given event\footnote{In our notation, a scalar product 
of $f$ and $g$
in the $L^2({\mathbb{R}}^2)$ space is denoted as $\langle f,g \rangle$:
$\langle f,g \rangle = \int \int f(x,y) g(x,y) \,dx \,dy$.
Repeated indices are being summed over, even if the summation symbol is
not written explicitly.}.

Using completeness of the basis, we expand
\begin{equation}
\rho - \bar{\rho} =
\langle \rho,F^\lambda_{m,i,j}\rangle F^\lambda_{m,i,j}
- \langle \bar{\rho},F^\lambda_{m,i,j}\rangle F^\lambda_{m,i,j}
\end{equation}

Notice that  $\bar{\rho}$,
being constant
within the detector's
rectangular acceptance,
is orthogonal
to any  $F^\lambda_{m,i,j}$ with $m \ge 1$.
Due to the
orthonormality condition
$\langle F^\lambda_{m,i,j},F^{\lambda'}_{m',i',j'}\rangle =
\delta_{\lambda,\lambda'}\delta_{m,m'}\delta_{i,i'}\delta_{j,j'}$,
the $\rho - \bar{\rho}$ components for
different scales do not form cross-terms in the $\sigma^2$ sum,
and the sum
contains no cross-terms between $\rho$ and $\bar{\rho}$ for the four
observable scales.
Instead of a $\langle \rho, G_{m=5,i,j} \rangle$ set, the
Si detector energy amplitude array
-- its closest experimentally achievable approximation -- is used as the
DWT input.
We used the WAILI \cite{WAILI} software library to obtain the wavelet
decompositions.

The Fourier power spectrum of a random white noise field
is known to be independent of frequency \cite{N_Wiener}.
We are looking for dynamical textures in the data, and  therefore
would like to treat the random white noise case as a ``trivial'' one
to compare with.
Therefore it is interesting to reformulate this property for
wavelets, where scale plays  the same role as frequency in Fourier
analysis.

To do that, we link scales with frequencies, or in other words,
we must understand the frequency spectra of the wavelets.
The Fourier images of 1D
wavelet functions occupy a set of wave numbers
whose characteristic broadness grows with scale fineness $m$ as $2^m$;
$2^{2m}$ should be used in the 2D case.
Discrete wavelets of higher orders have better frequency localization
than the Haar wavelets.
Despite this advantage, we use Haar because only Haar allows
one  to say that the act of data
taking  with the (binned !) detector constitutes the first stage of the
wavelet transformation.

In 2D, we find it most informative to present the three modes of a power
spectrum
with different directions of sensitivity
$P^{\phi\eta}(m)$, $P^\phi(m)$, $P^\eta(m)$
separately.
We define the \emph{power spectrum} as
\begin{equation}
P^\lambda(m) =
\frac{1}{2^{2m}}\sum_{i,j}\langle \rho,F^\lambda_{m,i,j}\rangle^2 ,
\label{eq:P_m}
\end{equation}
where the denominator gives the meaning of spectral \emph{density}
to the observable.
So defined, the $P^\lambda(m)$ of a random white noise field is
independent of $m$.
\begin{figure}
\epsfxsize=8.6cm
\epsfbox{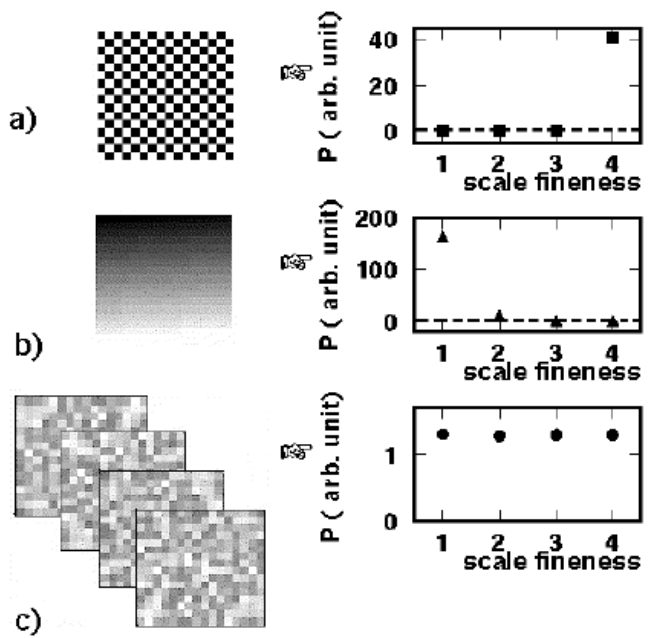}
\caption{
Understanding the analyzing potency of the DWT power spectra: a) for a checkerboard pattern
b) for a smooth gradient pattern c) for a sample of a thousand random white noise images -- 
in this case the \emph{average} power spectrum is shown.
}
\label{intuition}
\end{figure}

In order to illustrate the sensitivity of the wavelet transformation to texture features of
the different scales, we have applied the wavelet transform to three  test patterns, shown in
Fig. \ref{intuition}.
All patterns are $16\times16$ pixel matrices.
The left hand side shows the test pattern, and the right shows the power spectrum resulting
from the wavelet transform.
Pattern a), a checkerboard, has structure only on the finest scale 
and all power components of scales coarser than 4 are zero.
Pattern b) has exactly the opposite scale composition; the slow gradation between black
and white corresponds to a structure on the coarsest scale, as seen in the accompanying power
spectrum.  
Smoothness of the gradient means that neighbor-to-neighbor changes do not add much
to the pattern once the overall trend (the large scale feature) is taken into account.

These two examples illustrate the property of scale localization,
made possible by virtue of the
scale orthogonality of the basis.
Patterns encountered in multiple hadron production involve a variety of
scales, and yet they are more likely to be of type b), 
rather than a). 
An important conclusion follows immediately: 
in  this type of measurement, large acceptance, like the one used in this
analysis, rather than fine segmentation, is the way to accomplish  sensitivity.

Case c) shows patterns that arise from white noise.
They produce signals in the power spectrum independent of scale, as expected.
In the first approximation, the white noise example provides a
base-line case for comparisons in a search for non-trivial effects.
\begin{figure}
\epsfxsize=8.6cm
\epsfbox{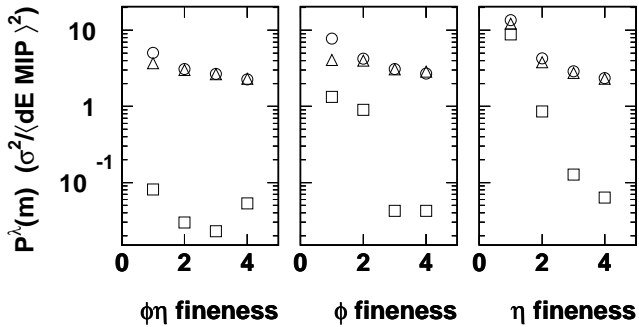}
\caption{
Power spectra of $7\times10^3$ events in the multiplicity bin
$326<\,dN/\,d\eta<398$ (between $\approx 6\%$ and 10\% centrality).
$\bigcirc$ -- true events,
$\bigtriangleup$ -- mixed events,
$\protect\Box$ -- the average event.
}
\label{compare}
\end{figure}

Figure \ref{compare}
shows the power spectra measured in $Pb+Pb$ for  one  multiplicity range.
The unit on the vertical scale ($\sigma^2/\langle dE_{MIP} \rangle^2$)
is chosen so that the power of fluctuations whose variance $\sigma^2$
equals the squared mean energy loss by a minimum ionizing particle
traversing the detector, is the unit.
The first striking feature is that the power spectra of physical events
are indeed enhanced on the coarse scale.
The task of the analysis is to quantify and, as much as possible, 
eliminate ``trivial'' and experiment-specific reasons for this 
enhancement.

\subsection{Identification and control of systematic errors}

The average event, formed by summing amplitude images of the
measured  events in a given multiplicity bin,
and dividing by the number of events, has a much reduced texture as
statistical fluctuations cancel (shown as $\protect\Box$ in 
Fig.\ref{compare}).
Average events retain the
texture associated with the shape of $\,d^2N/\,d\phi\,d\eta$, 
with the dead
channels and the finite  beam geometrical cross-section
(though this is only partially visible in the average event,
 due to the fact that event averaging is done
without attempting to select events according to the vertex position).
$P^\lambda(m)$ is proportional to the variance, or squared fluctuation
$\sigma^2$.
Therefore, for Poissonian statistics of hits in a pad, the event
averaging over $M$ events should  decrease $P^\lambda(m)$  by
a factor of $M$.
The average event whose power spectrum is shown on
Fig. \ref{compare} is formed by adding $7\times10^3$ events,
however its $P^\lambda(m)$ is down less than $7\times10^3$ compared to
that of the single events.
This demonstrates that the average event's texture is
not due to statistical fluctuations, but rather,
\emph{predominantly} due to the systematic uncertainties listed.
Consequently, we can use the average event's $P^\lambda(m)$ to estimate
the magnitude of the static texture-related systematics.
As seen from Fig. \ref{compare}, the systematics
are far below the $P^\lambda(m)$ of single events
(true or mixed),
with the exception of pseudorapidity, where
 non-constancy of $\,dN/\,d\eta$ over the detector's
acceptance is visible.

The way to get rid of the ``trivial'' or static texture is to use mixed
events, taking different channels from different events.
The mixed events preserve the texture associated with the
detector position offset,
the inherent $\,dN/\,d\eta$ shape and the dead channels.
This is \emph{static} texture as it produces
the same pattern event after event while
we are searching for evidence of dynamic texture.
We reduce sources of the static texture in the
power spectra by empty target subtraction and by subtraction of mixed
events power spectra, thus obtaining the \emph{dynamic texture}
$P^\lambda(m)_{true} - P^\lambda(m)_{mix}$.
In order to reproduce the electronics cross-talk effects
in the mixed event sample, the mixing is done sector-wise, i.e. the
sectors constitute
the subevents subjected to the event number scrambling.

We continue with a brief summary of the systematic errors in the 
measurements
of the DWT dynamic texture observable $P_{true} - P_{mix}$.
Static texture and dynamic background texture
present the largest problem in the search for the phase 
transition-related
dynamic texture via power spectra of local fluctuations.
The method of solving the problem is comparison with the reference 
sample
created by event mixing.
Thus the $P_{true} - P_{mix}$ observable was created.
For comparison with models, a Monte Carlo simulation of the Si detector
is used.
It includes the known static texture
effects and undergoes the same  procedure to remove the effects.
The ``irreducible remainder'' is the residual effect which may
\begin{enumerate}
\item{survive the elimination procedure}
\item{emerge \emph{as a difference} between the
data, subjected to the elimination procedure,
and the MC analyzed in the same manner.}
\end{enumerate}
Table \ref{tab:stat_texture} lists the sources of static texture and
summarizes the methods of their treatment.
We group the \emph{background texture} sources according to similarity 
of manifestation and treatment, into
\begin{itemize}
\item{statistical fluctuations}
\item{static texture}
\item{background dynamic texture}
\end{itemize}
The statistical fluctuation is  the most trivial item in this list.
Both event mixing (provided that mixing is done within
the proper multiplicity class) and MC comparison solve this problem.
The statistical fluctuations
do not result in irreducible systematic errors.

The static texture group includes:
\begin{itemize}
\item{geometrical offset of the detector with respect to the beam's
``center of gravity'' in the vertical plane
}
\item{dead pads
}
\item{$\,dN/\,d\eta$ shape -- a genuine large scale
multiparticle correlation sensitive to the physics of the
early stage of the collision
}
\end{itemize}
Cleanliness of the static texture elimination via event mixing has been
checked by simulating the contributing effects separately.
First, by running the detector response simulation on MC-generated 
events \emph{without} the beam/detector offset and with a
beam of 0 thickness it was ascertained that the remaining
dynamic texture is very small compared with the systematic errors
due to the background Si hits and the beam geometrical cross-section,
for all scales and all directional modes $\lambda$.
Due to the finite size of the multiplicity bin, the mixed events 
consist of subevents coming from events of different total multiplicity.
With the sector-wise mixing, this causes an additional sector-to-sector
variation of amplitude in the mixed events,
thus resulting in an enhancement of $P^{\phi}_{mix}$
primarily on the finest scale, with respect to $P^{\phi}_{true}$.
On Fig. \ref{multi_dep}, this effect can be seen as the
$P^{\phi}_{true}-P^{\phi}_{mix}$ values progressively
grow negative with multiplicity in the finest scale plot.
However, as can be seen on the same figure, the effect is small
compared with the total systematic error bars shown as boxes.

The background dynamic texture group includes:
\begin{itemize}
\item{elliptic and directed flow}
\item{finiteness of the beam cross-section}
\item{background hits in the Si}
\item{channel-to-channel cross-talk}
\end{itemize}

Elliptic and directed flow, observed at SPS \cite{NA49_elliptic},
are large scale dynamic texture phenomena of primarily azimuthal
(elliptic) and diagonal (directed flow) modes.
Because both reaction plane and direction angle vary event by event,
the respective dynamic textures
can not be subtracted by event mixing, unless the events are
classified according to their reaction plane orientation and
the direction angle, with mixing and $P_{true}-P_{mix}$
subtraction done within those classes.
Neither reaction plane nor direction angle was reconstructed
in the present analysis, and the $P_{true}-P_{mix}$
(especially that of the azimuthal and diagonal modes on the coarse
scale) retain the elliptic/directed flow contribution.
The effects of flow on dynamic texture observables
are smaller than other texture effects,
so they can not be singled out and quantified in this analysis.

The finite beam cross-section effect belongs to this group,
despite the fact that a very similar effect of geometrical 
detector/beam
offset has been classified as static texture.
An effect must survive mixing with its strength unaltered
in order to be fully subtracted via event mixing.
Preserving the effect of the random variations in the $Pb+Pb$
vertex on the power spectra
 in the mixed events requires classification of
  events according to the vertex position
and mixing only within such classes.
This requires knowledge of the vertex for each event,
which is not available in this experiment.
Therefore, MC simulation of the beam profile remains the only way to 
quantify false texture arising from vertex variations.
MC studies with event generators show that the
beam spatial extent and the resulting vertex variation is the source of
the growth
of the coarse scale \emph{azimuthal} texture correlation with 
multiplicity
(see Fig. \ref{multi_dep}).
Uncertainty in our knowledge of the beam's geometrical cross-section
must be propagated into a systematic error on $P_{true}-P_{mix}$.

The other two effects in this group are difficult to separate
and simulate and the error
estimate reflects the combined effect.
The systematic errors were evaluated by removing the $Pb$ target and
switching magnetic field polarity to expose the given side of the
detector
to $\delta$-electrons (from the air and T0),
 while minimizing nuclear interactions.
This gives an ``analog'' generator of uncorrelated noise.
All correlations (i.e. deviations of
$P^\lambda(m)_{true}$ from $P^\lambda(m)_{mix}$)
in this noise generator are treated as systematic uncertainties.
Thus this component of the systematic error gets a sign,
and the systematic errors are asymmetric.
The effect of increasing texture correlation (for diagonal and 
azimuthal modes)
with multiplicity on the coarse scale, attributed to the geometrical
offset of the detector with respect to the beam
(the leading one in the static group), is present in the switched
polarity empty target runs as well.
For this reason, it was impossible to disentangle the background 
dynamic
contribution on the coarsest scale.
In Table \ref{tab:stat_texture}, the ``irreducible remainder estimate''
for the diagonal, coarse scale is bracketed with two numbers, which 
form  the lower and upper estimates.
The lower estimate is obtained by taking the scale one unit finer and
quoting its number.
This, indeed, sets the lower limit because the deviations of
$P^\lambda(m)_{true}$ from $P^\lambda(m)_{mix}$
generally grow with scale coarseness.
The upper limit is set by ascribing the \emph{entire} texture
correlation,
observed in the $\delta$-electron data, to the background hits and 
channel
cross-talk, and ignoring the fact that significant portion of it must
be due
to the vertex fluctuation (finite beam profile).
This upper limit is likely to be a gross overestimation, and
in Fig. \ref{multi_dep} we show systematic errors, obtained by
adding in quadrature the finite beam error with the background hit 
error.

\section{Results}
\begin{figure}
\epsfxsize=8.6cm
\epsfbox{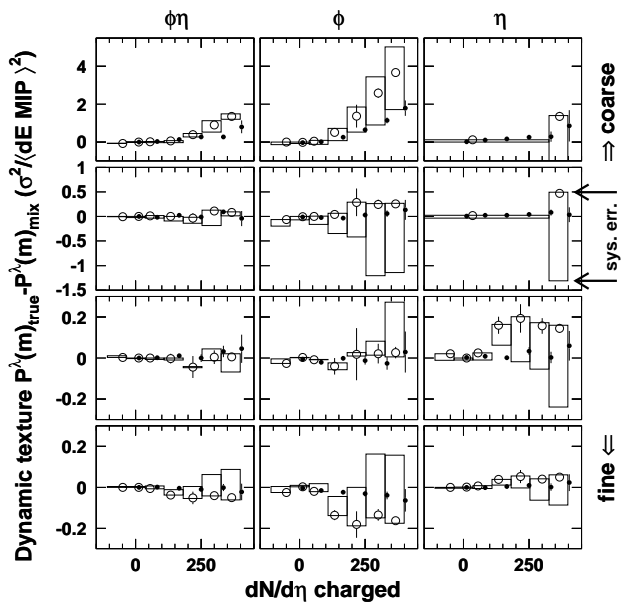}
\caption{
Multiplicity dependence of the texture correlation.
$\bigcirc$ -- the NA44 data, $\bullet$ -- RQMD.
The boxes show the systematic errors vertically and the boundaries of
the multiplicity bins horizontally; the statistical errors
are indicated by the vertical bars on the points. The rows correspond
to the scale fineness $m$, the columns -- to the directional mode
$\lambda$ (which can be diagonal $\phi\eta$,
azimuthal $\phi$, and pseudorapidity $\eta$).
        }
\label{multi_dep}
\end{figure}
Fig. \ref{multi_dep} presents a comparison of the DWT dynamic texture
in the measured and RQMD-simulated\cite{RQMD} $Pb+Pb$ collision events.
The three directional sensitivity modes
(diagonal $\phi\eta$, azimuthal $\phi$, and pseudorapidity $\eta$)
have four scales each, so that there are 12 sets of points
in the DWT dynamic texture as
a function of the charged multiplicity $\,dN_{ch}/\,d\eta$ bin.
The systematic errors on the points  (shown by vertical bars)
have been evaluated following the procedure described in detail in
Section
\ref{Section:analysis}.

\begin{figure}
\epsfxsize=8.6cm
\epsfbox{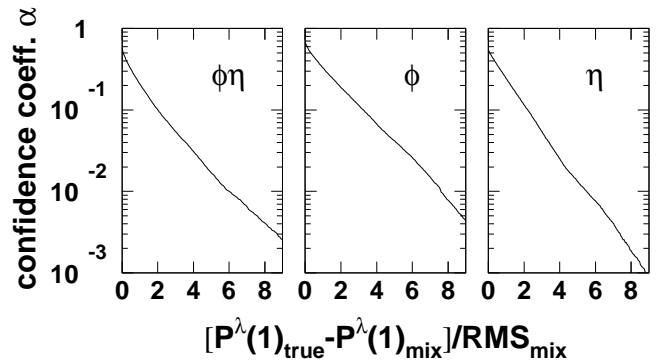}
\caption{Confidence coefficient
 as a function of the fluctuation strength.
 $RMS_{mix}$ denotes
 $\sqrt{\langle P^\lambda(1)_{mix}^2 -
 \langle P^\lambda(1)_{mix}\rangle^2\rangle}$.
The multiplicity
bin is $326<\,dN/\,d\eta<398$ (6-10\% centrality), as in Fig.\ref{compare}.
 }
\label{alpha_limit}
\end{figure}

Fig.\ref{compare}
demonstrated that the major fraction
of the observed texture exists also in mixed events.
A detailed account of the causes was discussed in the preceding 
section, including known physics as well as instrumental effects.
It is therefore clear that the observable most directly related to the
dynamical correlations/fluctuations is not $P^\lambda(m)$, but
$P^\lambda(m)_{true}-P^\lambda(m)_{mix}$.
This quantity, normalized to the $RMS$ fluctuation of 
$P^\lambda(m)_{mix}$,
can be used to characterize the relative strength of local fluctuations
in an event.
The distribution for different $\lambda$ (or directions)
is plotted on Figure \ref{alpha_limit}
in an integral way, i.e. as an $\alpha(x)$ graph where for every $x$,
$\alpha$ is the fraction of the distribution above $x$.
\begin{equation}
\alpha(x) =
{\int_x^{\infty} \frac{\,dN}{\,d\xi} \,d\xi}\Big/
{\int_{-\infty}^{+\infty} \frac{\,dN}{\,d\xi} \,d\xi},
\end{equation}
where $\xi$ denotes the
fluctuation strength
\begin{equation}
\xi = \frac{P^{\lambda}(1)_{true} - P^{\lambda}(1)_{mix}}
{RMS(P^{\lambda}(1)_{mix})},
\label{Eq:fluct_strength}
\end{equation}
and $\,dN/\,d\xi$ is the statistical distribution of $\xi$,
obtained from the experimentally known distributions of
$P^{\lambda}(1)_{true}$ and $P^{\lambda}(1)_{mix}$.
Expression \ref{Eq:fluct_strength} is constructed
 to be sensitive to the difference
between $P^{\lambda}(1)_{true}$ and $P^{\lambda}(1)_{mix}$, while
minimizing detector specifics to enable comparison between
different experiments in future.
The latter is accomplished by normalizing to $RMS_{mix}$.
This  normalization also eliminates the trivial multiplicity dependence
of the observable.

The fluctuation strength observable provides a
limit on the frequency and strength of the fluctuations and expresses
the result in a model-independent way.
The \emph{confidence level}
with which local fluctuations of a given strength (expressed through the
event by event observables via Eq. \ref{Eq:fluct_strength})
can be excluded
is then  $1-\alpha$.
Fluctuations greater than $3\times RMS_{mix}$ are excluded
in the azimuthal and pseudorapidity modes with 90\% and 95\% confidence,
respectively.
The monotonic fall of the curve is consistent with the absence of 
abnormal subsamples in the data.

RQMD events were fed into the GEANT detector response simulation
and analyzed using the same off-line
procedure as used for the experimental data.
The detector offset with respect to the beam center of gravity and
the beam profile were included in the simulation.
In a separate simulation run, the beam profile was identified as the
cause of the rise of the azimuthal dynamic texture with the
multiplicity on the coarse scale.
In our experiment, this purely instrumental effect dominates the
azimuthal component of the DWT dynamic texture.

The most apparent conclusion from Fig. \ref{multi_dep} is
that a large fraction of the texture (seen on Fig. \ref{compare}) is
not dynamic i.e. not different between true and mixed events.
Being monotonic (or absent), the change of the data points with 
multiplicity
does not reveal any evidence of a region of impact 
parameters/baryochemical
potentials with qualitatively different properties, such as those of
a critical point neighborhood.
The RQMD comparison confirms that
 particle production via hadronic multiple scattering, following string
 decays
(without critical phenomena or phase transition)
can explain the observed
results when detector imperfections are taken into account.
More detailed discussion of the implications of these data on various
phase transition models will be given in Section \ref{discussion}.

\section{Sensitivity}
\label{Si_DWT_sensitivity}
Interesting physics can manifest itself in the ensemble
probability density  distributions as well as in the
event-by-event (EbyE for short) observables.
To  illustrate  the power of the EbyE  observable we used,
we should construct   final  states of charged  particles  
indistinguishable
from the point of view of ``traditional'',  or ensemble-wise  
observables,
such as
\begin{enumerate}
\item{$\,dN/\,dy$ distribution}
\item{$\,dN/\,dp_T$, $1/mT \,dN/\,dm_T$ distribution etc.}
\item{ multiplicity distribution}
\end{enumerate}
and compare the sensitivity of the  above-mentioned  observables  with
that of the EbyE one.

A sensitivity study was performed using a multifireball event 
generator created specially for this purpose.
The generator produces textures of known magnitude
by simulating the observed multiplicity as arising from an arbitrary
number of fireballs.
Correlations among groups of particles arise when the particles
come from the same fireball.
We do not suggest that the physics of $Pb+Pb$ collisions is properly
described by a superposition of fireballs of a fixed size.
Rather, we use the fireballs as a way to generate controlled
multiparticle correlations.

This picture is inspired by Van Hove's scenario
\cite{VanHove:1985zy} of a first order phase transition via droplet 
fragmentation of a QGP fluid.
We measure texture in two directions, spanned by polar and azimuthal angles,
and are also sensitive to the spatial texture of
longitudinal flow.
For  boost-invariant expansion\cite{Bjorken} 
two droplets, separated along the longitudinal coordinate, will be
separated in $y$ and $\eta$.
As long as there is longitudinal expansion, a spatial texture will be
manifested as (pseudo)rapidity texture.
In the multifireball event generator, we generate the pseudorapidity 
texture explicitly, omitting the spatial  formulation of the problem.
The total $p_T$ of each fireball is 0;
its total $p_Z$ is chosen to reproduce the observed $\,dN/\,dy$
of charged particles by Lorentz-boosting the fireballs along the $Z$
direction, keeping the total $\vec{p}$ of an event at 0 in the rest
frame of the colliding nuclei.
The fireballs hadronize independently into charged and neutral pions
and kaons mixed in a realistic proportion.
By varying number of particles $N_p$ per fireball, one varies
``grain coarseness'' of the event texture in $\eta$.

\begin{figure}
\epsfxsize=8.6cm
\epsfbox{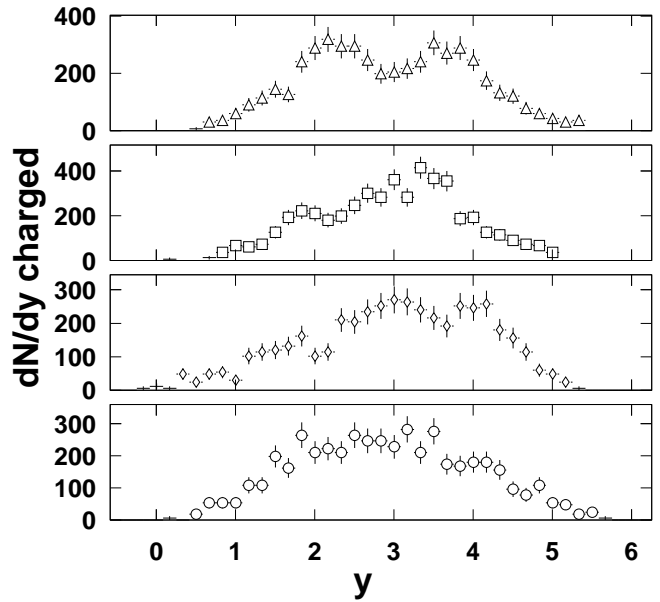}
\caption{$\,dN/\,dy$ distribution of charged particles in the 
multifireball
event generator in four \emph{individual events} with different number
of fireballs:
$\triangle$ -- 2 fireballs,
$\Box$ -- 4 fireballs,
$\Diamond$ -- 8 fireballs,
$\bigcirc$ -- 16 fireballs.
One can see how the texture becomes smoother as the number of fireballs
increases.
We remind  the reader that the detector's active area covers $\pi$
azimuthally and pseudorapidity 1.5 to 3.3.
In general, acceptance limitations make it more difficult to detect
dynamic textures.
 }
\label{texture_comparison}
\end{figure}
To illustrate the discussion,
Fig.\ref{texture_comparison} presents examples of $\,dN/\,dy$
distributions
in four events with different number of fireballs.
The dynamic textures seen on the figures are peculiar to these 
particular
events and are gone after $\,dN/\,dy$ of many events are added.

\begin{figure}[!th]
\epsfxsize=8.6cm
\epsfbox{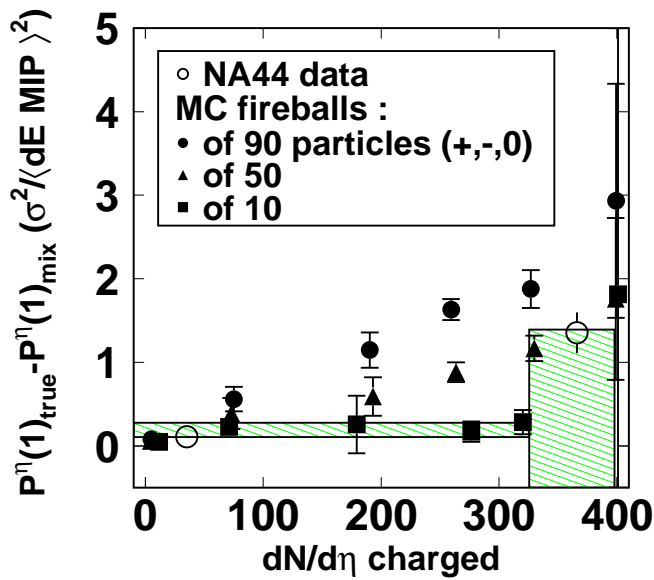}
\caption{
Coarse scale $\eta$ texture correlation in the NA44 data, shown by
$\bigcirc$ (from the top right plot of Figure \ref{multi_dep}),
is compared with that from the multifireball
event generator for three different fireball sizes.
Detector response is simulated.
The boxes represent systematic errorbars 
(see caption to Fig.~\ref{multi_dep}).
}
\label{sensitivity_press}
\end{figure}

We simulated average fireball multiplicities of
10, 50, 90 (with RMS fluctuation of 3) and larger.
Fig. \ref{sensitivity_press} shows comparison of our data with the 
simulated
 pseudorapidity texture.
With  $\sim 10^4$  events,  the
detector+software  can  differentiate between the cases of 50 and 90 particle 
fireballs.
The signal grows with the charged particle
multiplicity and with $N_p$.
Fig. \ref{sensitivity_press} provides \emph{quantitative} information
on the sensitivity of the texture measurements by relating the
expected strength of response to the strength of texture via Monte Carlo
simulation.
The sensitivity is limited by systematic errors of the measurement,
discussed in
Section \ref{Section:analysis}.
Nevertheless, it is instructive to compare
sensitivity of this method with other methods; in particular with
two point correlators.

The sensitivity of the method is remarkable indeed if one takes into 
account
that statistics in the fifth multiplicity bin for each of the three 
event
generator points is below $3\times10^4$ events -- too scarce, e.g., to
extract three source radius parameters via HBT analysis
even with a well optimized spectrometer!

The use of two particle correlation in rapidity
$R_2(y)$ to search
for droplets was discussed for  $p\bar{p}$ 
collisions at $\sqrt{s}=1.8$ TeV (at FNAL)\cite{Ruuskanen_Seibert}.
$R_2$ was reported  to \emph{decrease}
with multiplicity, so that it would not be expected to be visible
for $\,dN/\,dy$ above $\approx$20;
the signal would be \emph{weaker} in a scenario with correlated 
droplets.
In contrast,
the wavelet transformation retains sensitivity at high multiplicity, 
as we see in Fig.\ref{sensitivity_press}.
In the fifth multiplicity bin, with total number of hadrons at 
freeze-out
around $1.5\times10^2$, a typical fraction of particles coming 
from the same fireball
for the clustering parameters of 50 (90) would be 3\% (6\%)
\footnote{The fireball multiplicity has to be compared, of course, not with
 $\,dN/\,d\eta$, but with the
total multiplicity of the event, whose very rough estimate is
 $3/2\int\,d\eta \,dN_{ch}/\,d\eta = 3/2\int\,dy\,dN_{ch}/\,dy$, and if
one approximates $\,dN_{ch}/\,dy$ with a Gaussian whose 
$\sigma$-parameter is 1.4
(based on \cite{NA49_h-}),
the latter integral is $\approx 5 \max({\,dN/\,d\eta})$.}.
In either case there is little hope of seeing any trace of such
dynamics either in ensemble-averaged $\,dN/dy$ or in $\,dN/dy$ of a
single event, but the systematic difference between the power spectra of 
the real and mixed events, integrated over multiple events, nevertheless
reveals the difference.
The data are consistent with clustering among $\le$ 3\% of the particles.

\section{Discussion}
\label{discussion}

The order of the expected QCD phase transitions 
is known to be a complex issue for realistic current masses of quarks
 in the system of a finite size.
It is generally expected that a first order phase transition would be
easier to observe.
Our dynamic texture measurement  tests  the hypothesis of the first order
phase transition via
\emph{QGP droplet hadronization}\cite{VanHove:1985zy}
in a way more direct than  interpretation of $p_T$ spectra involving
latent heat.
Our result can be used to constrain
phenomenological quantities which represent basic QCD properties
and affect texture formation in this
class of hadronization models 
\cite{VanHove:1985zy,Mardor_Svetitsky,Cleymans_94}.
Such quantities are
the energy flux, or rate at which the QGP transmits its energy
to hadrons\cite{Banerjee,Muller_Eisenberg},
 critical size of the QGP droplet\cite{Mardor_Svetitsky}, and
initial upper energy density of the transition $\epsilon_0^\prime$.

The specific experimental signature of second order phase transition 
(known since the
discovery of critical opalescence \cite{opalescence}) is the 
emergence of
critical fluctuations of the order parameter with an enormous 
increase of the correlation lengths.
However, for physical quark masses Rajagopal and Wilczek
\cite{Rajagopal_Wilczek,Rajagopal:1993ah}
argued that due to closeness of the pion mass to the critical 
temperature,
it would be unlikely for the correlation volumes to include large 
numbers of
pions, if the cooling of the plasma and hadronization proceeds in an
equilibrated manner.
If, on the contrary, the high temperature configuration \emph{suddenly}
finds itself
at a low temperature, a self-organized criticality regime settles in,
and the critical local fluctuations
develop fully\cite{Rajagopal_Wilczek,Rajagopal:1993ah}.

The NA44 data reported here signifies absence of dynamical fluctuations 
on the scales probed,
within the limit of sensitivity discussed in Section 
\ref{Si_DWT_sensitivity}.
Convincing evidence of thermal equilibration can be provided
by event-by-event observables.
Our data is consistent with local thermal equilibrium,
understood as an absence of physically distinguished
scales between the scale of
a hadron and the scale of the system, or scale invariance of 
fluctuations \cite{Trainor:2000dm} (``white noise'').
However to \emph{probe} equilibration \emph{directly}
with this method, texture sensitivity at least down to the typical
fireball (cluster) sizes observed in $pN$ collisions in cosmic rays and
accelerator experiments \cite{Cocconi,Drijard} would be necessary.
In the absence of such direct evidence, the
non-observation of critical fluctuations can imply either absence of
the second order phase transition or presence of thermal equilibration -- the 
latter voids the criticality signature, according to 
Rajagopal and Wilczek
\cite{Rajagopal_Wilczek}.



\section{Conclusion}
We have developed a method of measuring the dynamic component of local
fluctuations in charged particle density in pseudorapidity and azimuthal
angle, and applied the analysis to $Pb+Pb$ collisions
measured by the NA44 experiment.
Comparison of the data to a 
simple Monte Carlo texture event generator indicates that 
sensitivity to pseudorapidity density clusters of $\ge 3\%$  
is accomplished in this experiment.
The probability of encountering a real event whose dynamic azimuthal texture
exceeds in strength that of a random mixed event by 3 \emph{RMS}, 
is below 10\%.
For the pseudorapidity texture, the respective probability is below 5\%. 

We conclude that this novel method of event-by-event analysis,
sensitive to particular
signatures of first and second order phase transitions,
does not reveal such signatures in 158 GeV/A
$Pb+Pb$ collisions at the SPS.


\section{Acknowledgement}
The authors thank N.~Antoniou, I.~Dremin, K.~Rajagopal,
 E.~Shuryak, M.~Stephanov,
and T.~Trainor for illuminating discussions.
The NA44 Collaboration thanks the staff of the CERN PS-SPS
accelerator complex for their excellent work, and the technical staff
in the collaborating institutes for their valuable contributions.
This work was supported by the Austrian Fonds zur F{\"o}rderung der
Wissenschaftlichen Forschung;
the Science Research Council of Denmark;
the Japanese Society for the Promotion of Science; the Ministry of
Education, Science and Culture, Japan;  the Science Research Council
of Sweden; the US Department of Energy and the National Science
 Foundation.

\begin{table}
\begin{ruledtabular}
\begin{tabular}{cccccc}
             & \multicolumn{4}{c}{Correction}               & Irredu-\\
\cline{2-5}
            &           & \multicolumn{2}{c}{event mixing} &  & cible\\
\cline{3-4}
 Source     & subtract & subtract    & preserve & do      & remain-  \\
            & empty   & mixed       & sectors  & MC      &  der  \\
            & target  & events      &          &         &   estimate\\
     &        &        &    &  &$\sigma^2/\langle dE_{MIP} \rangle^2$\\
\hline
stat. fluctuations     & N/A&  yes     & N/A & yes  & 0. \\
\hline
$\,dN/\,d\eta$ shape, &N/A  & yes       & OK       &  yes &  0. \\
offset, dead pads     &     &           &          &      &  \\
\hline
finite beam         &       &       &      & yes & $0.14$ \\
Xsection            & N/A   &   N/A & N/A  &     &        \\
$1\times2$ mm       &       &       &      &     &        \\
\hline
background hits      &  yes    & yes    & yes  & can't  & \\
\cline{1-5}
channel Xtalk    &     &      &      &      & \Bigg\}$>0.070,<0.37$  \\
8.5\% for            &  N/A    &  yes   & yes  & can't  &        \\
neighbors           &         &        &      &        &        \\
\end{tabular}
\caption{Sources of background texture (dynamic and static) and 
their treatment.
The irreducible remainder estimate is quoted for diagonal texture 
correlation
in the $326< \,dN/\,d\eta <398$ bin, and is expressed in the units of
$\sigma^2/\langle dE_{MIP} \rangle^2$;
see text for information on how it was obtained.}
\label{tab:stat_texture}
\end{ruledtabular}
\end{table}



\begin{thebibliography}{99}
\bibitem{EMU15}
N.~M.~Astafeva, I.~M.~Dremin and K.~A.~Kotelnikov,
Mod.\ Phys.\ Lett.\ A {\bf 12} 1185-1192 (1997);
I.~M.~Dremin, O.~V.~Ivanov, S.~A.~Kalinin, K.~A.~Kotelnikov, 
V.~A.~Nechitailo and N.~G.~Polukhina,
Phys.\ Lett.\ B {\bf 499}, 97 (2001)
[hep-ph/0007060].
\bibitem{NA49_phi}
H.~Appelshauser {\it et al.}  [NA49 Collaboration],
Phys.\ Lett.\ B {\bf 459}, 679 (1999)
[hep-ex/9904014].
\bibitem{broken_ergo}
R.~G.~Palmer,
Adv.\ Phys.\  {\bf 31}, 669 (1982).
\bibitem{VanHove:1985zy}
L.~Van~Hove,
Z.\ Phys.\ C {\bf 27}, 135 (1985).
\bibitem{deflagration}
L.~Van~Hove,
Z.\ Phys.\ C {\bf 21}, 93 (1983).
\bibitem{Mardor_Svetitsky}
I.~Mardor and B.~Svetitsky, Phys.\ Rev.\ D {\bf 44},  878 (1991).
\bibitem{Cleymans_93}
N.~Bilic, J.~Cleymans, E.~Suhonen and D.~W.~von Oertzen,
 Phys.\ Lett.\ B {\bf 311}, 266 (1993).
\bibitem{Cleymans_94}
N.~Bilic, J.~Cleymans, K.~Redlich and E.~Suhonen,
Z.\ Phys.\ C {\bf 63}, 525 (1994)
[arXiv:hep-ph/9307351].
\bibitem{Svetitsky_Yaffe}
B.~Svetitsky and L.~Yaffe,
Nucl.\ Phys.\  {\bf B210}, 423 (1982).
\bibitem{Pisarski_Wilczek}
R.~Pisarski and F.~Wilczek, Phys.\ Rep.\ D {\bf 29}, 338 (1989).
\bibitem{FRBrown}
F.~R.~Brown, F.~P.~Butler, H.~Chen, N.~H.~Christ, Z.~Dong, W.~Schaffer, 
L.~I.~Unger, and A.~Vaccarino, Phys.\ Rev.\ Lett.\ {\bf 65}, 2491 (1990).
\bibitem{Rajagopal_Wilczek}
K.~Rajagopal and F.~Wilczek, Nucl.\ Phys.\ {\bf B399}, 395 (1993).
\bibitem{tricritical}
M.~Stephanov, K.~Rajagopal and E.~Shuryak,
Phys.\ Rev.\ Lett.\ {\bf 81}, 4816 (1998).
\bibitem{Shuryak:2001pd}
E.~V.~Shuryak and M.~A.~Stephanov,
Phys.\ Rev.\ C {\bf 63}, 064903 (2001).
[hep-ph/0010100].
\bibitem{flow_AGS}
J.~Barrette {\it et al.}  [E877 Collaboration],
Phys.\ Rev.\ C {\bf 55}, 1420 (1997);
C.~Pinkenburg {\it et al.}  [E895 Collaboration],
Phys.\ Rev.\ Lett.\  {\bf 83}, 1295 (1999)
[nucl-ex/9903010].
\bibitem{NA49_elliptic}
H.~Appelshauser {\it et al.}
NA49 Collaboration,
Phys.\ Rev.\ Lett.\ {\bf 80}, 4136 (1998).
[nucl-ex/9711001].
\bibitem{DWT}
I.~Daubechies, \emph{Ten Lectures on Wavelets}
(SIAM, Philadelphia, 1992) and references therein, p.129.
\bibitem{NA44ex}
H.~B{\o}ggild {\it et al.}, [NA44 Collaboration],
Phys.\ Lett.\ B {\bf 302}, 510 (1993)
[Erratum-ibid.\ B {\bf 306}, 418 (1993)].
\bibitem{AMPLEX}
E.~Beuville, K.~Borer, E.~Chesi, E.~H.~Heijne, P.~Jarron, B.~Lisowski 
and S.~Singh,
Nucl.\ Instrum.\ Meth.\ A {\bf 288}, 157 (1990).
\bibitem{Landau_distr}
L.~D.~Landau, J.~Phys.~USSR, {\bf 8}, 201 (1944).
\bibitem{MINUIT}
F.~James. MINUIT. Function Minimization and Error Analysis.
CERN Program Library Long Writeup D506
(CN Division, CERN, 1994).
\bibitem{DWT_power}
J.~Pando and L.~Fang, Phys.\ Rev.\ E{\bf 57}, 3593-3601 (1998);
L.~Fang and J.~Pando,
astro-ph/9701228.
\bibitem{WAILI}
 G.~Uytterhoeven \emph{et al.},
 WAILI: Wavelets with Integer Lifting.
 TW Report 262, Department of Computer Science,
 Katholieke Universiteit Leuven, Belgium, July 1997.
\bibitem{N_Wiener}
N. Wiener, \emph{Generalized Harmonic Analysis, and Tauberian Theorems}
(Cambridge, Mass., M.I.T. Press 1966), p.116.
\bibitem{RQMD}
H.~Sorge, Phys.\ Rev.\ C {\bf 52}  3291 (1995);
we use version 2.4 of RQMD.
\bibitem{Bjorken}
J.~D.~Bjorken, Phys.\ Rev.\ D {\bf 27}, No.1  140 (1983).
\bibitem{Ruuskanen_Seibert}
P.~V.~Ruuskanen, D.~Seibert, Phys.\ Lett.\ B {\bf 213}, 227 (1988).
\bibitem{NA49_h-}
P.~G.~Jones for the NA49 Collaboration, Nucl.\ Phys.\ {\bf A610}, 188c (1996).
\bibitem{Banerjee}
B.~Banerjee, N.~K.~Glendenning and T.~Matsui,
Phys.\ Lett.\ B {\bf 127}, 453 (1983).
\bibitem{Muller_Eisenberg}
B.~M\"{u}ller and J.~M.~Eisenberg, Nucl.\ Phys.\  {\bf A435}, 
791 (1985).
\bibitem{opalescence}
M.~Altschul, Zs.\ phys.\ Chem., {\bf 11}, 578 (1893);
K.~von~Wesendonck, Naturwiss.\ Rundschau, {\bf 9}, 210 (1894).
\bibitem{Rajagopal:1993ah}
K.~Rajagopal and F.~Wilczek,
Nucl.\ Phys.\  {\bf B404}, 577 (1993)
[hep-ph/9303281].
\bibitem{Trainor:2000dm}
T.~A.~Trainor,
hep-ph/0001148.
\bibitem{Cocconi}
G.~Cocconi, Phys. Rev. {\bf 111}, 1699 (1958).
\bibitem{Drijard}
D.~Drijard \emph{et al.}, Nucl. Phys. {\bf B155},  269 (1979).
\end{thebibliography}
\end{document}